# Revisiting the First, Second and Combined Laws of Thermodynamics

Zi-Kui Liu

Department of Materials Science and Engineering, The Pennsylvania State University,

University Park, Pennsylvania 16802, USA

Chen and Mauro [1] published a paper interpreting the first law of thermodynamics for open systems and commented on the present author's definition of the first law of thermodynamics [2–7] (Eq. 2 in their paper, Eq. 13 in the last paper by the present author [6] among the four papers cited in their Endnote 4) and chemical potential (Eq. 3 in their paper, Eq. 14 in ref. [6]). The present author would like to thank them for pointing out the error in Eq. 13 and Eq. 14 in ref. [6]. Accordingly, with Eq. 13 in ref. [6] containing the term of $-PdV$, Eq. 14 in ref. [6] should be revised as follows

$$\mu_i = U_i - TS_i + PV_i \qquad Eq.\ 1$$

where the definitions of all variables in Eq. 1 and the rest of the present paper are shown in the footnotes of Table 1 and Table 2, respectively.

This issue concerns fundamentally how the first, second, and combined laws of thermodynamics are written in textbooks and taught in classrooms and is worth further discussion. These laws address three distinctive aspects of a system [8]. The first law concerns the exchanges of heat, work, and mass between a system and its surroundings. The second law focuses on internal processes and the admissible direction of these processes. The combination of the first and second laws is based on the definition of the entropy change of the system and results in the combined law of thermodynamics. The partial derivative of internal energy with respect to a molar quantity results in various potentials, including $U_i$, $T$, $-P$, and $\mu_i$ as shown in Table 1, noting that "potential theory involves problems describable in terms of a partial differential equation, in which the dependent variable is the appropriate potential" [9]. These four partial derivatives, i.e., potentials, are related to each other by Eq. 1.

From Table 1, it seems that one would have $dW$ equal to $-PdV$ from the first law by Gibbs [10,11] and Hillert [8]. However, this is only valid for a closed system in the first law by Gibbs, i.e., $dN_i = 0$. The two equations in the first law by Hillert shown in Table 1, i.e., Eqs. 1.7 and 1.11 in his book [8], imply that $dW$ is equal to $-PdV$ for an open system, which is incorrect. The present author followed Hillert's first law (column Liu_2016 in Table 1 [12]) and corrected it in his next paper (column Liu_2020 in Table 1 [2]) by introducing the partial internal energy of a component, $U_i$. However, the present author did not realize that when the hydrostatic pressure is introduced, the total work exchange between a system and its surroundings should be defined as follows

$$dW = -P\left(dV - \sum V_i dN_i\right) \qquad Eq.\ 2$$

because the exchange of mass between a system and its surroundings affects the volume of the system as shown by Eq. 14 in the paper by Chen and Mauro [1] and Eq. 2 in the present paper. Consequently, the $PV_i$ term must be added to the definition of the chemical potential in Eq. 14 in ref. [6] as shown by Eq. 1 in the present paper.

It is noted in Table 1 that both Gibbs [10,11] and Hillert [8] introduced the term of $-PdV$ into the first law before entropy was introduced. For an open system, i.e., $dN \neq 0$, Hillert introduced $H_m$ without using Eq. 2 above, resulting in the error in the two forms of the first law as shown in Table 1 and Eqs. 1.7 and 1.11 in his book [8]. Berry et al. [13] considered the hydrostatic work and the mass addition in two steps. The first step is with $dQ = 0$ and $dW = 0$, i.e., $dV = \sum V_i dN_i$ from Eq. 2. The second step is to compress the system, so its volume is not changed due to the mass addition, i.e., $dV = 0$ and $dW = P \sum V_i dN_i$ from Eq. 2. The first law of thermodynamics under $dQ = 0$ and $dV = 0$ is thus written as follows [13]

$$dU = dW + \sum U_i dN_i = -P\left(dV - \sum V_i dN_i\right) + \sum U_i dN_i = \sum (U_i + PV_i) dN_i \quad Eq.\ 3$$

which was what Hillert [8] obtained. When $dQ \neq 0$ and $dV \neq 0$, the first law of thermodynamics becomes

$$dU = dQ - PdV + \sum (U_i + PV_i) dN_i \quad Eq.\ 4$$

It is important to emphasize that $-PdV \neq dW$ in Eq. 4, and their relation is shown by Eq. 2 above for hydrostatic work, which was the origin of the error in Eq. 14 in ref. [6].

In the other three papers [2,3,5] cited in the Endnote 4 of the paper by Chen and Mauro [1], the present author developed the combined law of thermodynamics by combining the first and second laws without specifying the types of work (column Liu_2020 in Table 1) and defined the chemical potential as $U_i - TS_i$. However, this definition is inconsistent with the definition of chemical potential by Gibbs and Hillert who introduced the hydrostatic work in the first law of thermodynamics. Therefore, chemical potential should be defined using Eq. 1 in the present paper when $-PdV$ is used in the combined law.

Consequently, the chemical potential for different types of work exchange between a system and its surroundings are different due to their unique contributions from mass exchange, as shown in Table 2, i.e., hydrostatic or nonhydrostatic mechanical works, electric work, and magnetic work, respectively [14]. The last row in Table 2 represents the most general form of the chemical potential and the combined law of thermodynamics. Furthermore, the first law defined by Chen and Mauro [1], i.e., their Eq. 7 for equilibrium systems and Eq. 45 for non-equilibrium closed systems, where $d_{ip}S$ was introduced through an *ad hoc* process with different definitions of $dS$ in their Eq. 7 and Eq. 45, can be related to the Gibb's and Hillert's combined laws through the definition of $dS$ shown in Table 1.

It is noted that the present author [3,6,15] divided the entropy production of each independent internal process into four actions: (1) heat generation $(d_{ip}Q)$, (2) consumption of some components as reactants $(dN_{r,j})$ with entropy $S_j$, (3) formation of some components as products $(dN_{p,k})$ with entropy $S_k$, and (4) reorganization of its configurations $(d_{ip}S^{config})$, as follows,

$$d_{ip}S = \frac{d_{ip}Q}{T} - \sum_j S_j dN_{r,j} + \sum_k S_k dN_{p,k} + d_{ip}S^{config} \geq 0 \qquad Eq.\ 5$$

It was pointed out that the microscopic 'violation of second law of thermodynamics' discussed in the literature is due to the incomplete consideration of all contributions to entropy production [3,5,6].

**Acknowledgements**

The present author thanks Long-Qing Chen and John Mauro for pointing out and discussing the error in chemical potential when hydrostatic work was specified in the combined law of thermodynamics. The author thanks Xiaofeng Guo, John Mauro, David McDowell, and Yu Zhong for their valuable comments on the manuscript. This work was supported by the U.S. Department of Energy (DOE) through Grant No. DE-SC0023185 and the Endowed Dorothy Pate Enright Professorship at Penn State.

*Table 1: First, second, and combined laws of thermodynamics and their derivations by Gibbs [10,11], Hillert [8], and the present author [2,12]. The red and **bold** texts, i.e., $\mathbf{H_m}$ and $\mathbf{H_i}$, are incorrect and should be replaced by $U_m$ and $U_i$, respectively.*

| | Gibbs [10,11] | Hillert [8] | Liu_2016 [12] | Liu_2020 [2] |
|---|---|---|---|---|
| First Law, $dU$ | $dQ + dW$ $= dQ - PdV$ | $dQ + dW + \mathbf{H_m} dN$ $= dQ - PdV + H_m dN$ | $dQ + dW + \sum \mathbf{H_i} dN_i$ | $dQ + dW + \sum U_i dN_i$ |
| Second Law | $d_{ip}S = 0$ | \multicolumn{3}{c}{$d_{ip}S \geq 0$} | | |
| Entropy Change, $dS$ | $\dfrac{dQ}{T}$ | $\dfrac{dQ}{T} + S_m dN + d_{ip}S$ | \multicolumn{2}{c}{$\dfrac{dQ}{T} + \sum S_i dN_i + d_{ip}S$} | |
| Combined Law for Multicomponent Systems, $dU$ | $TdS - PdV$ $+ \sum \mu_i dN_i$ | $TdS - PdV + \sum \mu_i dN_i - Td_{ip}S$ $= \sum Y^a dX^a - \sum D_j d\xi_j$ | | $TdS + dW$ $+ \sum (U_i - TS_i) dN_i$ $- Td_{ip}S$ |

- $dU$: Internal energy change in the system
- $dQ$: Exchange of heat between the system and its surroundings
- $P$: Pressure
- $V$: Volume
- $dW$: Exchange of work between the system and its surroundings
- $U_m$: Molar internal energy of the system
- $H_m$: Molar enthalpy of the system
- $N$: Moles of the system
- $N_i$: Moles of component $i$
- $U_i = \left(\dfrac{\partial U}{\partial N_i}\right)_{dQ=0, dW=0, N_{j \neq i}}$: Partial internal energy of component $i$

- $d_{ip}S$: Entropy production due to internal processes ($ip$)

- $dS$: Total entropy change of the system

- $S_m$: Molar entropy of the system

- $S_i = \left(\frac{\partial S}{\partial N_i}\right)_{dQ=0, N_{j\neq i}, d_{ip}S=0}$ : Partial entropy of component $i$

- $\mu_i = \left(\frac{\partial U}{\partial N_i}\right)_{S,V,N_{j\neq i}, d_{ip}S=0}$ : Chemical potential of component $i$

- $Y^a = \left(\frac{\partial U}{\partial X^a}\right)_{X^{b\neq a}, d_{ip}S=0}$ : Potentials, denoting $T$, $-P$, and $\mu_i$ in the combined law

- $X^a$: Conjugate molar quantities of $Y^a$, i.e., $S$, $V$, and $N_i$, in the combined law

- $\xi_j$: Internal variables, represented by an independent internal molar quantity

- $D_j = \left(\frac{\partial U}{\partial \xi_j}\right)_{S,V,N_i}$ : Driving force for the internal process ($d\xi_j$), represented by the difference of the conjugate potential of $\xi_j$ due to $d\xi_j$

- $Td_{ip}S = \sum D_j d\xi_j$: The overall contribution of independent internal processes to the total entropy production.

*Table 2: Chemical potential and combined laws of thermodynamics under various types of work*

| Work | $dW$ | Chemical Potential of component $i$, $\mu_i$ | Combined Law of Thermodynamics, $dU$ |
|---|---|---|---|
| Hydrostatic | $-P\left(dV - \sum V_i dN_i\right)$ | $U_i - TS_i + PV_i$ | $TdS - PdV + \sum \mu_i dN_i - Td_{ip}S$ |
| Mechanical | $-V\sigma\left(d\varepsilon - \sum \varepsilon_i dN_i\right)$ | $U_i - TS_i + V\sigma\varepsilon_i$ | $TdS - V\sigma d\varepsilon + \sum \mu_i dN_i - Td_{ip}S$ |
| Electric | $-VE\left(d\theta - \sum \theta_i dN_i\right)$ | $U_i - TS_i + VE\theta_i$ | $TdS - VEd\theta + \sum \mu_i dN_i - Td_{ip}S$ |
| Magnetic | $-V\mathcal{H}\left(dB - \sum B_i dN_i\right)$ | $U_i - TS_i + V\mathcal{H}B_i$ | $TdS - V\mathcal{H}dB + \sum \mu_i dN_i - Td_{ip}S$ |
| Mechanical + Electric + Magnetic | $-V\left[\sigma\left(d\varepsilon - \sum \varepsilon_i dN_i\right) + E\left(d\theta - \sum \theta_i dN_i\right) + \mathcal{H}\left(dB - \sum B_i dN_i\right)\right]$ | $U_i - TS_i + V(\sigma\varepsilon_i + E\theta_i + \mathcal{H}B_i)$ | $TdS - V(\sigma d\varepsilon + Ed\theta + \mathcal{H}dB) + \sum \mu_i dN_i - Td_{ip}S$ |

- $V_i = \left(\frac{\partial V}{\partial N_i}\right)_{dW=0, N_{j\neq i}}$ : Partial volume of component $i$

- $\sigma$ and $\varepsilon$: Stress and strain tensors

- $\varepsilon_i = \left(\frac{\partial \varepsilon}{\partial N_i}\right)_{dW=0, N_{j\neq i}}$ : Partial strain tensor of component $i$

- $E$ and $\theta$: Electric field and electric displacement tensors

- $\theta_i = \left(\frac{\partial \theta}{\partial N_i}\right)_{dW=0, N_{j\neq i}}$ : Partial electric displacement tensor of component $i$

- $\mathcal{H}$ and $B$: Magnetic field and magnetic flux tensors

- $B_i = \left(\frac{\partial B}{\partial N_i}\right)_{dW=0, N_{j\neq i}}$ : Partial magnetic flux tensor of component $i$